\documentclass[prb,showpacs,twocolumn]{revtex4}%
\usepackage{amsfonts}
\usepackage{amsmath}
\usepackage{amssymb}
\usepackage{graphicx}%
\begin{document}
\title{Orbital magnetization and its effects in spin-chiral ferromagnetic
\textit{kagom\'{e}} lattice}
\author{Zhigang Wang and Ping Zhang}
\affiliation{Institute of Applied Physics and Computational Mathematics, P.O. Box 8009,
Beijing 100088, P.R. China}

\begin{abstract}
Recently, Berry phase in the semiclassical dynamical of Bloch electrons has
been found to make a correction to the phase-space density of states and a
general multi-band formula for finite-temperature orbital magnetization has
been given [Phys. Rev. Lett. \textbf{97}, 026603 (2006)], where the orbital
magnetization $\mathcal{M}$ consists of two parts, i.e., the conventional part
$M_{c}$ and the Berry-phase correction part $M_{\Omega}$. Using this general
formula, we theoretically investigate the orbital magnetization and its
effects on thermoelectric transport and magnetic susceptibility properties of
the two-dimensional \textit{kagom\'{e}} lattice with spin anisotropies
included. The study in this paper is highly interesting by the occurrence of
nonzero Chern number in the lattice. The spin chirality parameter $\phi$ (see
text) results in profound effects on the orbital magnetization properties. It
is found that the two parts in orbital magnetization opposite each other. In
particular, we show that $M_{c}$ and $M_{\Omega}$ yield the paramagnetic and
diamagnetic responses, respectively. It is further shown that the orbital
magnetization displays fully different behavior in the metallic and insulating
regions, which is due to the different roles $M_{c}$ and $M_{\Omega}$ play in
these two regions. The anomalous Nernst conductivity is also calculated, which
displays a peak-valley structure as a function of the electron Fermi energy.

\end{abstract}
\maketitle

\section{Introduction}

In the semiclassical dynamics of Bloch electrons, \ the Liouville's theorem on
the conservation of phase-space volume is violated by the occurrence of Berry
phase, which leads to a modification of the phase-space density of states
\cite{Xiao1}. This modified phase-space density of states, $D_{n}%
(\mathbf{r},\mathbf{k})$=$1$+$e\mathbf{B}{\small \cdot}\mathbf{\Omega}%
_{n}/\hbar$, enters naturally in the semiclassical expression for the
expectation value of physical quantities, and has profound effects on
equilibrium as well as transport properties. Here $\mathbf{B}$ is the external
magnetic field, and $\mathbf{\Omega}_{n}(\mathbf{k})$ is the Berry curvature
(gauge flux) of electronic Bloch states defined by $\mathbf{\Omega}%
_{n}(\mathbf{k})=i\langle\nabla_{\mathbf{k}}u_{n}(\mathbf{k})|\times
|\nabla_{\mathbf{k}}u_{n}(\mathbf{k})\rangle$ with $|u_{n}(\mathbf{k})\rangle$
being the periodic part of Bloch wave for the $n$-th band. Based on this
correction in the semiclassical phase-space density, Xiao et al. \cite{Xiao2}
have further derived a general expression for the single-particle free energy
in the presence of the external magnetic field as follows (for a general three
dimensional system)
\begin{equation}
F=-\frac{1}{\beta}\sum_{n}\int d^{3}\mathbf{k}\left(  1+\frac{e}{\hbar
}\mathbf{B}\cdot\mathbf{\Omega}_{n}(\mathbf{k})\right)  \ln[1+e^{\beta
(\mu-\varepsilon_{n}^{M})}]. \tag{1}%
\end{equation}
Here $\mu$ is the electron chemical potential, $\beta=1/k_{B}T$, and
$\varepsilon_{n}^{M}=\varepsilon_{n\mathbf{k}}-\mathbf{m}_{n}(\mathbf{k}%
)\cdot\mathbf{B}$ is the electron band energy in the presence of the external
magnetic field. $\mathbf{m}_{n}(\mathbf{k})$ is the crystal orbital magnetic
moment and conventionally defined by \cite{Chang,Sund} $\mathbf{m}%
_{n}(\mathbf{k})=-i(e/2\hbar)\langle\nabla_{\mathbf{k}}u_{n}(\mathbf{k}%
)|\times\lbrack\hat{H}(\mathbf{k})-\varepsilon_{n}(\mathbf{k})]|\nabla
_{\mathbf{k}}u_{n}(\mathbf{k})\rangle$, where $\hat{H}(\mathbf{k})$ is the
crystal Hamiltonian acting on $|u_{n}(\mathbf{k})\rangle$. The orbital
magnetization (OM) is then given by the field derivative at fixed temperature
and chemical potential, $\mathcal{M}(\mathbf{r})=-\left(  \partial
F/\partial\mathbf{B}\right)  _{\mu,T}$, with the result%
\begin{align}
\mathcal{M}(\mathbf{r})  &  =\sum_{n}\int d^{3}\mathbf{km}_{n}(\mathbf{k}%
)f_{n}(\mathbf{r},\mathbf{k})\tag{2}\\
&  +\frac{1}{\beta}\sum_{n}\int d^{3}\mathbf{k}\frac{e}{\hbar}\mathbf{\Omega
}_{n}\ln\left[  1+e^{\beta(\mu-\varepsilon_{n}^{M})}\right] \nonumber\\
&  \equiv\mathbf{M}_{c}+\mathbf{M}_{\mathbf{\Omega}},\nonumber
\end{align}
where $f_{n}(\mathbf{r},\mathbf{k})$ is the local equilibrium Fermi-Dirac
distribution function for $n$-th band. The first term in Eq. (2) is just a
statistical sum of the orbital magnetic moments of the carriers originating
from the self-rotation of the carrier wave packet \cite{Chang,Sund}, thus we
call this term the conventional part $M_{c}$ of the OM. Whereas the second
term $M_{\Omega}$ is the Berry phase correction to the OM. This term is of
topological nature, arising from a bulk consideration. Interestingly, it is
this Berry-phase term that eventually enters the transport current
\cite{Xiao2}. At zero temperature and magnetic field the general expression
(2) reduces to \cite{Xiao1,Thon1}%
\begin{equation}
\mathcal{M}=\sum_{n}\int^{\mu_{0}}d^{3}\mathbf{k}\left(  \mathbf{m}%
_{n}(\mathbf{k})+\frac{e}{\hbar}\mathbf{\Omega}_{n}(\mathbf{k})\left[  \mu
_{0}-\varepsilon_{n\mathbf{k}}\right]  \right)  , \tag{3}%
\end{equation}
where the upper limit means that the integral is over states with energies
below the Fermi energy $\mu_{0}$. As mentioned above, the first term in Eq.
(3) is the contribution from the intrinsic orbital moment, and the second term
comes from the Berry-phase correction of the density of states.

Among very few studies on Berry phase effect of the OM, Xiao et al.
\cite{Xiao2} have investigated the anomalous thermoelectric transport in
CuCr$_{2}$Se$_{4-x}$Br$_{x}$ system \cite{Lee} by use of their derived
relation between anomalous Nernst effect (ANE) and anomalous Hall effect
(AHE). Ceresoli et al. \cite{Thon2} have given an extensive tight-binding
calculation of OM for finite and periodic two-dimensional (2D) Haldane model
\cite{Haldane1} in a wide range of parameters covering cases of Chern number
(a dimensionless integer \cite{Thouless}) $C=0$ and $C\neq0$. Due to its basic
importance in understanding the magnetism and transport features of the
materials, obviously, more work are needed in exploiting the properties of the
OM in various kinds of realistic physical systems. Also the theory itself,
such as a full quantum mechanical derivation of the OM, remains to be further developed.

In this paper we extend the study of the OM to the strongly correlated
electronic systems. More specificially, we focus our attention to a typical
flat-band ferromagnet with spin anisotropies on the 2D \textit{kagom\'{e}}
lattice. This attention is partially motivated by the recently established
common point that the spin Berry phase plays an important role in the quantum
transport in spin-orbit coupled \cite{Jung,Fang,Yao} or spin-chiral
\cite{Matl,Chun,Ye,Tag} ferromagnetic systems; the latter is exampled by
pyrochlore compounds R$_{2}$Mo$_{2}$O$_{7}$ (R$=$Nd, Sm, Gd), in which the
spin configuration is noncoplanar and the spin chirality appears. As a
consequence, the quantum transport of electrons, especially the transverse
conductivity $\sigma_{xy}$ is expected to be affected by the presence of spin
chirality. In fact, recent transport experiments on ferromagnetic pyrochlores
have revealed that the AHE increases as the temperature $T$ is lowered and
approaches to the saturated value \cite{Taguchi, Katsufuji}. This behavior is
intrinsically different from the conventional theories \cite{Hurd}. One
explanation to this anomalous feature is that the pyrochlore structure has
geometrical frustration \cite{Ramirez} which consists of corner-sharing
tetrahedrons. Thus the antiferromagnetic and even the ferromagnetic
\cite{Harris} interaction between nearest-neighbor spins are frustrated.
Recently, Ohgushi et al. \cite{Ohgushi} have first pointed out that the chiral
spin state can be realized by the introduction of spin anisotropy in an
\textit{ordered} spin system on the 2D \textit{kagom\'{e}} lattice, which is
the cross section of the pyrochlore lattice perpendicular to the $(1,1,1)$
direction \cite{Ramirez}. In this case, it has been shown \cite{Ohgushi} that
the presence of chiral spin state may induce nonzero Chern number, thus
resulting in a quantized Hall effect in insulating state, which is expected to
have important implications to AHE experiments in ferromagnetic pyrochlores.

Motivated by the above work \cite{Ohgushi} on quantized Hall effect in the 2D
\textit{kagom\'{e}} lattice, in this paper we turn to study the OM properties
and effects in this 2D lattice system with nonzero Chern number. We show that
the OM displays different behavior in the metallic and insulating regions,
which is due to the different roles $M_{c}$ and $M_{\Omega}$ play in these two
regions. Also, the ANE and the orbital magnetic susceptibility are
investigated, which further illustrate the fundamental role brought forth by
the Berry-phase contribution and nonzero Chern number.

\section{Theoretical model and Chern number analysis}

Now we consider the double-exchange ferromagnet on the \textit{kagom\'{e}}
lattice \cite{Ohgushi} schematically shown in Fig. 1. Here the triangle is the
one face of the tetrahedron, and the easy axis of the spin anisotropy points
to the center of each tetrahedron and has an out-of-plane component. In this
situation the three local spins on sites A, B, and C in Fig. 1 have different
directions and the spin chirality emerges. The effective Hamiltonian for the
hopping electrons strongly Hund-coupled to these localized spins is given by
$H=\sum_{NN}t_{ij}^{eff}\psi_{i}^{\dag}\psi_{j}$ with $t_{ij}^{eff}%
=t\langle\chi_{i}|\chi_{j}\rangle=te^{ia_{ij}}\cos\frac{\vartheta_{ij}}{2}$.
Here the spin wave function $|\chi_{i}\rangle$ is explicitly given by
$|\chi_{i}\rangle=\left[  \cos\frac{\vartheta_{i}}{2},\text{ }e^{i\phi_{i}%
}\sin\frac{\vartheta_{i}}{2}\right]  ^{\text{T}}$, where the polar coordinates
are pinned by the local spins, i.e., $\langle\chi_{i}|\mathbf{S}_{i}|\chi
_{i}\rangle=\frac{1}{2}\left(  \sin\vartheta_{i}\cos\phi_{i},\text{ }%
\sin\vartheta_{i}\sin\phi_{i},\text{ }\cos\vartheta_{i}\right)  $.
$\vartheta_{ij}$ is the angle between the two spins $\mathbf{S}_{i}$ and
$\mathbf{S}_{j}$. The phase factor $a_{ij}$ can be regarded as the gauge
vector potential $a_{\mu}(\mathbf{r})$, and the corresponding gauge flux is
related to scalar spin chirality $\chi_{ijk}$=$\mathbf{S}_{i}{\small \cdot
}(\mathbf{S}_{j}{\small \times}\mathbf{S}_{k})$ \cite{Laughlin}. In periodic
crystal lattices, the nonvanishing of the gauge flux relies on the multiband
structure with each band being characterized by a Chern number
\cite{Thouless,Shindou}. Here the Chern number appears as a result of the spin
chirality in ferromagnets. Following Ref. \cite{Ohgushi} we set the flux
originated from the spin chirality per triangle (see Fig. 1) as $\phi$, which
satisfies $e^{i\phi}=e^{i(a_{AB}+a_{BC}+a_{CA})}$. The flux penetrating one
hexagon in Fig. 1 is determined as $-2\phi$. We take the gauge, in which the
phase of $t_{ij}^{eff}$ is the same for all the nearest-neighbor pairs with
the direction shown by the arrows in Fig. 1. It should be pointed out that the
net flux through a unit cell vanishes due to the cancelation of the
contribution of the two triangles and a hexagon. Also noted is that the
time-reversal symmetry is broken except for cases of $\phi$=$0$,$\pi$.%

\begin{figure}[tbp]
\begin{center}
\includegraphics[width=1.0\linewidth]{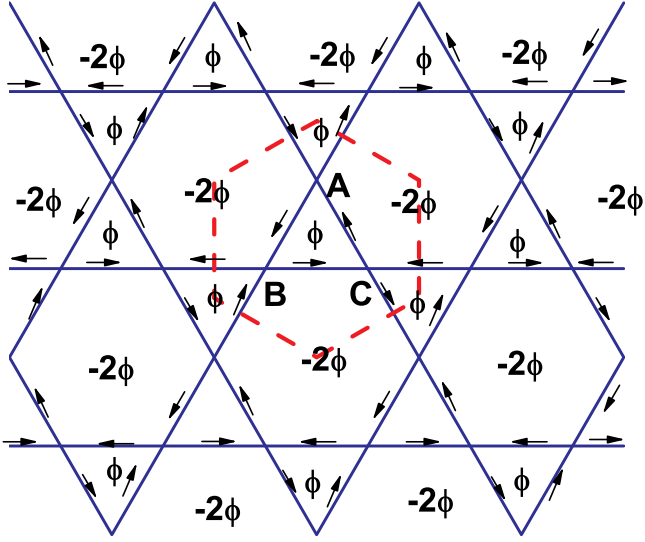}
\end{center}
\caption
{(Color online). Two dimensional spin-chiral ferromagnetic
\textit{kagom\'{e}} lattice. The dashed line represents the Wigner-Seitz unit
cell, which contains three independent sites (A, B, C). It is assumed that
each site has a different spin anisotropy axis. The arrows means the sign of
the phase of the transfer integral $t_{ij}$.}
\label{fig1}
\end{figure}%
The momentum-transformation of the above tight-binding Hamiltonian is given by
\cite{Ohgushi} \begin{widetext}%
\begin{equation}
H(\mathbf{k})=\left(
\begin{array}
[c]{ccc}%
0 & 2\cos(\mathbf{k}\cdot\mathbf{a}_{1})e^{-i\phi/3} & 2\cos(\mathbf{k}%
\cdot\mathbf{a}_{3})e^{i\phi/3}\\
2\cos(\mathbf{k}\cdot\mathbf{a}_{1})e^{i\phi/3} & 0 & 2\cos(\mathbf{k}%
\cdot\mathbf{a}_{2})e^{-i\phi/3}\\
2\cos(\mathbf{k}\cdot\mathbf{a}_{3})e^{-i\phi/3} & 2\cos(\mathbf{k}%
\cdot\mathbf{a}_{2})e^{i\phi/3} & 0
\end{array}
\right)  , \tag{4}%
\end{equation}
\end{widetext}where $\mathbf{a}_{1}$=$(-1/2,-\sqrt{3}/2)$, $\mathbf{a}_{2}%
$=$(1,0)$, and $\mathbf{a}_{3}=(-1/2,\sqrt{3}/2)$ represent the displacements
in a unit cell from A to B site, from B to C site, and from C to A site
respectively. In this notation, the Brillouin zone (BZ) is a hexagon with the
corners of $\mathbf{k}=\pm(2\pi/3)\mathbf{a}_{1}$, $\pm(2\pi/3)\mathbf{a}_{2}%
$, $\pm(2\pi/3)\mathbf{a}_{3}$, two of which are independent. Note that in
writing the Hamiltonian (4) we have chosen the unit of $t\cos(\vartheta
_{ij}/2)=1$, and set the length of each bond as unity. The eigenenergies of
the Hamiltonian (4) are given by
\begin{align}
\varepsilon_{1\mathbf{k}} &  =4\sqrt{\frac{b(\mathbf{k})}{3}}\cos\frac
{\theta(\mathbf{k})+2\pi}{3},\nonumber\\
\varepsilon_{2\mathbf{k}} &  =4\sqrt{\frac{b(\mathbf{k})}{3}}\cos\frac
{\theta(\mathbf{k})-2\pi}{3},\tag{5}\\
\varepsilon_{3\mathbf{k}} &  =4\sqrt{\frac{b(\mathbf{k})}{3}}\cos\frac
{\theta(\mathbf{k})}{3},\nonumber
\end{align}
where $b(\mathbf{k})$=$\cos^{2}(\mathbf{k}\cdot\mathbf{a}_{1})$+$\cos
^{2}(\mathbf{k}\cdot\mathbf{a}_{2})$+$\cos^{2}(\mathbf{k}\cdot\mathbf{a}_{3}%
)$, $\theta(\mathbf{k})$=$\arg\left[  \sqrt{\frac{27}{b^{3}(\mathbf{k})}%
}c(\mathbf{k})\text{+}i\sqrt{1-\frac{27c^{2}(\mathbf{k})}{b^{3}(\mathbf{k})}%
}\right]  $, and $c(\mathbf{k})$=$\cos(\mathbf{k}\cdot\mathbf{a}_{1}%
)\cos(\mathbf{k}\cdot\mathbf{a}_{2})\cos(\mathbf{k}\cdot\mathbf{a}_{3}%
)\cos\phi$. Because the relation $\varepsilon_{1\mathbf{k}}\leq\varepsilon
_{2\mathbf{k}}\leq\varepsilon_{3\mathbf{k}}$ is always satisfied, henceforth
we will call these energy bands `lower', `middle', and `upper' bands
respectively. The corresponding eigenvectors are \begin{widetext}%
\begin{equation}
|u_{n\mathbf{k}}\rangle=G_{n}(\mathbf{k})\left(
\begin{array}
[c]{c}%
\frac{1}{2}\left[  \varepsilon_{n\mathbf{k}}^{2}-4\cos^{2}(\mathbf{k}%
\cdot\mathbf{a}_{2})\right] \\
e^{i\frac{\phi}{3}}\left[  \varepsilon_{n\mathbf{k}}\cos(\mathbf{k}%
\cdot\mathbf{a}_{1})+2\cos(\mathbf{k}\cdot\mathbf{a}_{2})\cos(\mathbf{k}%
\cdot\mathbf{a}_{3})e^{-i\phi}\right] \\
e^{-i\frac{\phi}{3}}\left[  \varepsilon_{n\mathbf{k}}\cos(\mathbf{k}%
\cdot\mathbf{a}_{3})+2\cos(\mathbf{k}\cdot\mathbf{a}_{1})\cos(\mathbf{k}%
\cdot\mathbf{a}_{2})e^{-i\phi}\right]
\end{array}
\right)  , \tag{6}%
\end{equation}
\end{widetext}where the normalized factors are given by\begin{widetext}
\begin{equation}
G_{n}(\mathbf{k})=\frac{1}{\sqrt{2b(\mathbf{k})\varepsilon_{n\mathbf{k}}%
^{2}+(4b(\mathbf{k})-3\varepsilon_{n\mathbf{k}}^{2})\cos^{2}(\mathbf{k}%
\cdot\mathbf{a}_{2})+12c(\mathbf{k})\varepsilon_{n\mathbf{k}}}}. \tag{7}%
\end{equation}
\end{widetext}

First let us see the Chern number and Hall conductivity of this system, which
has been briefly reported in Ref. \cite{Ohgushi}. It is clear that the Hall
conductivity $\sigma_{xy}$ is equal to zero in the time-reversal symmetric
cases $\phi=0,$ $\pi$. Therefore, we focus on the case of $\phi\neq0,$ $\pi$.
In this situation there is an energy gap between the two nearest-neighbor
bands. We assume that the Fermi energy is lying in the gap. Then the Hall
conductivity is a sum over occupied Bloch bands, $\sigma_{xy}$=$(e^{2}%
/h)\sum_{n}^{\text{occu}}C_{n}$, where the $n$-th band Chern number is defined
by
\begin{align}
C_{n} &  =-\frac{1}{2\pi}\int_{\text{BZ}}d^{2}\mathbf{k}\Omega_{n}%
(\mathbf{k})\tag{8}\\
&  =-\frac{1}{2\pi}\int_{\text{BZ}}d^{2}\mathbf{k}\hat{z}\cdot\left(
\nabla_{\mathbf{k}}\times\mathbf{A}_{n}(\mathbf{k})\right)  ,\nonumber
\end{align}
where $\mathbf{A}_{n}(\mathbf{k})$=$i\langle u_{n}(\mathbf{k})|\nabla
_{\mathbf{k}}u_{n}(\mathbf{k})\rangle$ is the Berry phase connection (vector
potential) for the $n$-th band. According to the expressions for
$|u_{n\mathbf{k}}\rangle$, one obtains the expression for $\mathbf{A}%
_{n}(\mathbf{k})$ as follows
\begin{equation}
\mathbf{A}_{n}(\mathbf{k})=2\varepsilon_{n\mathbf{k}}G_{n}^{2}(\mathbf{k}%
)\sin\phi\cos k_{x}\left[  \sin\sqrt{3}k_{y}\mathbf{\hat{x}}+\sqrt{3}\sin
k_{x}\mathbf{\hat{y}}\right]  .\tag{9}%
\end{equation}
To proceed with Eqs. (8)-(9), one may first transform the integral of the curl
of the vector potential $\mathbf{A}_{n}$ over BZ to the line intergal of
$\mathbf{A}_{n}$ along the BZ boundary by use of Stokes' theorem, and then
apply the complex contour integration technique and residue theorem to
sinusoidal functions. After a straightforward derivation, one obtains
$C_{1}=-$sgn$(\sin\phi)$, $C_{2}=0$, and $C_{3}=$sgn$(\sin\phi)$,
respectively, which means that the quantum Hall effect is realized. However,
this purely mathematical calculation of Chern number is not favored by
theoretical physicists, who instead resort to the physical connotation that
the vector potential $\mathbf{A}_{n}$ and gauge flux $\mathbf{\Omega}_{n}$ are
endowed with. Correspondingly, we start this gauge-field analysis with the
notation that the value of Chern number is invariant under gauge
transformation $|u_{n\mathbf{k}}^{\prime}\rangle$=$e^{ig_{n}(\mathbf{k}%
)}|u_{n\mathbf{k}}\rangle$, $\mathbf{A}_{n}^{\prime}(\mathbf{k})=\mathbf{A}%
_{n}(\mathbf{k})-\nabla_{\mathbf{k}}g_{n}(\mathbf{k})$, where $g_{n}%
(\mathbf{k})$ is an arbitrary smooth function of $\mathbf{k}$. The gauge of
the wave functions Eq. (6) has been chosen in such way that the first
component is real. If this gauge choice is applicable in the whole region of
the BZ, the Chern number will obviously be zero. However, at two equivalent BZ
edge points $\mathbf{k}_{0}=(0,\pm\pi/\sqrt{3})$ on BZ boundary, one can find
that the wave functions $|u_{n\mathbf{k}}\rangle$ in Eq. (6) is ill-defined
since both its denominator and numerator are zero at these two points. This
means that the used gauge cannot apply to the whole BZ and one needs to render
a gauge transformation to avoid the singularity at $\mathbf{k}_{0}$. For this
one writes down the other set of eigenvectors as follows \begin{widetext}
\begin{equation}
|u_{n\mathbf{k}}^{\prime}\rangle=G_{n}^{\prime}(\mathbf{k})\left(
\begin{array}
[c]{c}%
e^{-i\frac{\phi}{3}}\left[  \varepsilon_{n\mathbf{k}}\cos(\mathbf{k}%
\cdot\mathbf{a}_{1})+2\cos(\mathbf{k}\cdot\mathbf{a}_{2})\cos(\mathbf{k}%
\cdot\mathbf{a}_{3})e^{i\phi}\right] \\
\frac{1}{2}\left[  \varepsilon_{n\mathbf{k}}^{2}-4\cos^{2}(\mathbf{k}%
\cdot\mathbf{a}_{3})\right] \\
e^{i\frac{\phi}{3}}\left[  \varepsilon_{n\mathbf{k}}\cos(\mathbf{k}%
\cdot\mathbf{a}_{2})+2\cos(\mathbf{k}\cdot\mathbf{a}_{1})\cos(\mathbf{k}%
\cdot\mathbf{a}_{3})e^{i\phi}\right]
\end{array}
\right)  , \tag{10}%
\end{equation}
\end{widetext}where the second component turns to be real. The normalized
factors are given by \begin{widetext}
\begin{equation}
G_{n}^{\prime}(\mathbf{k})=\frac{1}{\sqrt{2b(\mathbf{k})\varepsilon
_{n\mathbf{k}}^{2}+(4b(\mathbf{k})-3\varepsilon_{n\mathbf{k}}^{2})\cos
^{2}(\mathbf{k}\cdot\mathbf{a}_{3})+12c(\mathbf{k})\varepsilon_{n\mathbf{k}}}%
}. \tag{11}%
\end{equation}
\end{widetext}The new eigenvectors $|u_{n\mathbf{k}}^{\prime}\rangle$ recovers
the well-defined behavior at $\mathbf{k}_{0}$; the new singular points brought
force by $|u_{n\mathbf{k}}^{\prime}\rangle$ are $\mathbf{k}_{0}^{\prime}%
=\pm(\frac{\pi}{2},\frac{\pi}{2\sqrt{3}})$. Thus according to the two
different gauge choices, the BZ is now divided into two regions V and
V$^{\prime}$ as shown in Fig. 2. The wave functions $|u_{n\mathbf{k}}\rangle$
are used onto the region V, while $|u_{n\mathbf{k}}^{\prime}\rangle$ apply to
V$^{\prime}$. Note that there remains some freedom in the division of the BZ.
Because $|u_{n\mathbf{k}}\rangle$ and $|u_{n\mathbf{k}}^{\prime}\rangle$ are
ill-defined only at $\mathbf{k}_{0}$ and $\mathbf{k}_{0}^{\prime}$,
respectively, we are free to deform this division as long as $\mathbf{k}_{0}%
$($\mathbf{k}_{0}^{\prime}$)$\notin$V(V$^{\prime}$). This corresponds to the
gauge degree of freedom \cite{Muk2003}. At \textbf{k}$\mathbf{\in}$V$\cap
$V$^{\prime}$, the two choices of wave functions are different by a phase
factor $|u_{n\mathbf{k}}^{\prime}\rangle$=$e^{ig_{n}(\mathbf{k})}%
|u_{n\mathbf{k}}\rangle$, i.e., $\mathbf{A}_{n}^{\prime}(\mathbf{k}%
)$=$\mathbf{A}_{n}(\mathbf{k})-\nabla_{\mathbf{k}}g_{n}(\mathbf{k})$, where
\begin{widetext}
\begin{equation}
e^{ig_{n}(\mathbf{k})}=\frac{G_{n}^{\prime}(\mathbf{k})}{G_{n}(\mathbf{k}%
)}\frac{e^{-i\frac{\phi}{3}}\left[  \varepsilon_{n\mathbf{k}}\cos
(\mathbf{k}\cdot\mathbf{a}_{1})+2\cos(\mathbf{k}\cdot\mathbf{a}_{2}%
)\cos(\mathbf{k}\cdot\mathbf{a}_{3})e^{i\phi}\right]  }{\frac{1}{2}\left[
\varepsilon_{n\mathbf{k}}^{2}-4\cos^{2}(\mathbf{k}\cdot\mathbf{a}_{2})\right]
}. \tag{12}%
\end{equation}
\end{widetext}Thus one obtains the values of nonzero Chern number for lower
and upper bands as follows
\begin{align}
C_{n} &  =-\frac{1}{2\pi}\oint_{\partial\text{V}}\left[  \mathbf{A}%
_{n}(\mathbf{k})-\mathbf{A}_{n}^{\prime}(\mathbf{k})\right]  {\small \cdot
}d\mathbf{k}=-\frac{1}{2\pi}\oint_{\partial\text{V}}dg_{n}(\mathbf{k}%
)\tag{13}\\
&  =\left\{
\begin{array}
[c]{l}%
-\text{sgn}(\sin\phi)\text{ \ \ (lower band)}\\
\text{sgn}(\sin\phi)\text{ \ \ \ \ (upper band)}%
\end{array}
\right.  ,\nonumber
\end{align}
which is confirmed by the explicit calculation based on the complex-contour
integration technique. The nonzero Chern number implies that the wave function
cannot be written as a single function for the entire BZ. This also affects
the definition of field operators $a_{n\mathbf{k}}$ \cite{Muk2003}. Let
$a_{n\mathbf{k}}$ denote the annihilation operators when \textbf{k}$\in$V.
Then $|u_{n\mathbf{k}}^{\prime}\rangle$=$e^{ig_{n}(\mathbf{k})}|u_{n\mathbf{k}%
}\rangle$ implies the field operator to be $e^{-ig_{n}(\mathbf{k}%
)}a_{n\mathbf{k}}$ at \textbf{k}$\in$V$^{\prime}$. This phase mismatch between
patches in the BZ provides the quantized Hall conductivity.%
\begin{figure}[tbp]
\begin{center}
\includegraphics[width=1.0\linewidth]{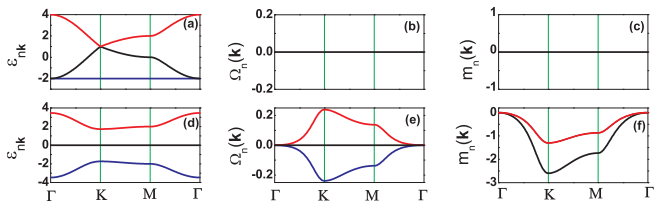}
\end{center}
\caption{(Color online). Division of the Brillouin zone of the \textit
{kagom\'{e}}
lattice model into two regions V (red area) and V$^{\prime}$ (blue area).}
\label{fig2}
\end{figure}
\

\section{Orbital Magnetization and its effects}

Now we turn to study the OM and its various effects. In the present
two-dimensional case, the magnetization and Berry curvature become
pseudoscalar quantities $M_{c}$, $M_{\Omega}$, and $\Omega_{n}$. From Eq. (3)
one can see that for band insulator there will be a discontinuity in OM if the
integral of the Berry curvature over the entire BZ, or the anomalous Hall
conductivity, is nonzero and quantized. The size of the discontinuity is given
by the quantized anomalous Hall conductivity times $E_{g}/e$, where $E_{g}$ is
the energy gap. Because an analytic derivation of the OM for 2D
\textit{kagom\'{e}} lattice is very tedious, thus we turn instead to a full
numerical representation in terms of the eigenstates $|u_{n\mathbf{k}}\rangle$
and eigenenergies $\varepsilon_{n\mathbf{k}}$ given in Eqs. (5)-(6).%

\begin{figure}[tbp]
\begin{center}
\includegraphics[width=1.0\linewidth]{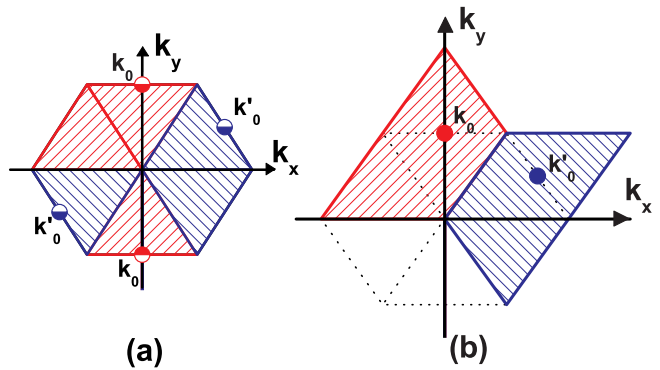}
\end{center}
\caption{(Color online). (From left to right) band strcture, Berry curvature
$\Omega_{n}(\mathbf{k})$, and conventional orbital magnetic moment
$m_{n}(\mathbf{k})$. The blue, black, and red curves correspond to the lower,
middel, and upper bands respectively. The spin chirality parameter is chosen
to be $\phi$=$0$ for upper panels while $\phi$=$\pi/2$ for
lower panels.}
\label{fig3}
\end{figure}%
We first show in Figs. 3 structures of the energy bands $\varepsilon
_{n\mathbf{k}}$, the crystal orbital moment $m_{n}(\mathbf{k})$, and the Berry
curvature $\Omega_{n}(\mathbf{k})$ along high-symmetry lines in the BZ for the
values of $\phi$=$0$ (upper panels) and $\phi$=$\pi/2$ (lower panels). In the
case of $\phi$=$0$, the time-reversal symmetry is preserved. From Fig. 3(a)
one can see that for $\phi$=$0$ the middle band and upper band touch at the
corner points (denoted by $\mathbf{k}_{\text{K}}$) in the BZ, around which the
upper and middle bands exhibit a cusp, $\epsilon_{3\mathbf{k}}$ ($\epsilon
_{2\mathbf{k}}$)=1$\pm\sqrt{3}|\mathbf{k}-\mathbf{k}_{\text{K}}|$. In this
case, the low-energy quasiparticle excitations can be well understood within a
(2+1)-D Dirac fermion field theory. The lower band becomes dispersionless
($\varepsilon_{1\mathbf{k}}$=$-2$) at $\phi$=$0$, which reflects the fact that
the 2D \textit{kagom\'{e}} lattice is a line graph of the honeycomb structure
\cite{Mie}. This flat band touches at the $\Gamma$ point of the BZ with the
middle band, whose dispersion around $\Gamma$ looks like an isotropic
paraboloid, $\varepsilon_{2\mathbf{k}}$=$-2$+$\frac{1}{8}k^{2}$. Note that the
energy spectrum for $\phi$=$\pi$ (not shown here) is a particle-hole conjugate
of that for $\phi$=$0$; therefore, the upper band becomes flat with an
eigenvalue of 2. The Berry curvature $\Omega_{n}(\mathbf{k})$ and the
intrinsic orbital magnetic moment $m_{n}(\mathbf{k})$ for three bands are zero
everywhere in the whole BZ in the time-reversal symmetric case of $\phi$=$0$,
as shown in Figs. 3(b)-(c). This can be simply seen by the fact that the wave
function $|u_{n\mathbf{k}}\rangle$ for $\phi$=$0$ is real. Generally they
satisfy the property $m_{n}(-\mathbf{k})$=$-m_{n}(\mathbf{k})$, $\Omega
_{n}(-\mathbf{k})$=$-\Omega_{n}(\mathbf{k})$ under time-reversal symmetry and
$m_{n}(-\mathbf{k})$=$m_{n}(\mathbf{k})$, $\Omega_{n}(-\mathbf{k})$%
=$\Omega_{n}(\mathbf{k})$ under spatial inversion symmetry. Thus in a crystal
with both time-reversal symmetry and spatial inversion symmetry,
$\mathbf{\Omega}_{n}(\mathbf{k})$ and $\mathbf{m}_{n}(\mathbf{k})$ will
disappear in the BZ. Figures 3(d)-(f) show $\varepsilon_{n\mathbf{k}}$,
$\Omega_{n}(\mathbf{k})$, and $m_{n}(\mathbf{k})$ for the value of $\phi$%
=$\pi/2$. In this case the time-reversal symmetry is broken. One can see that
there does no longer exist degeneracy in the energy spectrum and two gaps
among three bands now open. The middle band turns to become flat for $\phi
$=$\pi/2$, and the energy spectra shows a particle-hole symmetry. Note that
generally the energy spectra has no particle-hole symmetry except for the
cases of $\phi$=$\pm\pi/2$. This particle-hole symmetry at $\phi$=$\pi/2$ is
also reflected in $\Omega_{n}(\mathbf{k})$ [Fig. 3(e)] through the feature
that the Berry curvature for the flat middle band is zero everywhere in the
BZ, while the Berry curvatures for lower and upper bands are nonzero and
different with a sign. The intrinsic magnetic moment $m_{n}(\mathbf{k})$ for
$\phi$=$\pi/2$ is shown in Fig. 3(f). One can see that unlike the Berry
curvature, $m_{2}(\mathbf{k})$ for the middle flat band is nonzero in the BZ
and its amplitude is the sum of $m_{1}(\mathbf{k})$ and $m_{3}(\mathbf{k})$,
which are equal in the presence of particle-hole symmetry.%

\begin{figure}[tbp]
\begin{center}
\includegraphics[width=1.0\linewidth]{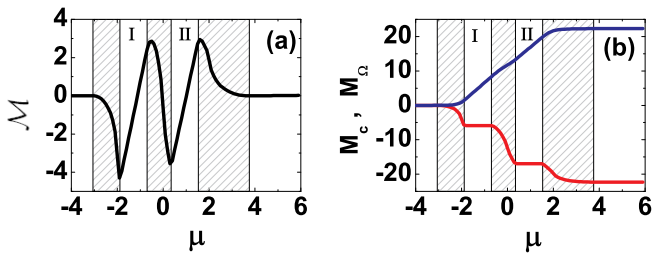}
\end{center}
\caption{(Color online). (a) Orbital magnetization of the \textit{kagom\'{e}}
lattice,
and (b) its two components $M_{c}$ (red curve) and $M_{\Omega}$ (blue curve)
as a function of the electron chemical potential $\mu$ for $\phi$=$\pi/3$. The
shaded areas correspond to the three groups of bands. To suppress the
divergence at band/gap contacts, we have used the temperature of $k_{B}
T$=$0.05$.}
\label{fig4}
\end{figure}%
Figure 4(a) shows the OM ($\mathcal{M}$)\ as a function of the electron
chemical potential $\mu$ for the value of $\phi$=$\pi/3$, in which case
neither time-reversal symmetry nor particle-hole symmetry is preserved. One
can see that initially the OM rapidly decreases as the filling of the lower
band increases, arriving at a minimum at $\mu$=$-2$, a value corresponding to
the top of the lower band. Then, as the chemical potential continues to vary
in the gap [region I in Fig. 4(a)] between the lower band and middle band, the
OM goes up and increases as a linear function of $\mu$. This linear
relationship in the insulating region can be understood by Eq. (3), from which
one obtains
\begin{align}
\frac{d\mathcal{M}}{d\mu} &  =\frac{e}{\hbar}\sum_{n}^{\text{occu}}\int
\frac{d^{2}k}{(2\pi)^{2}}\Omega_{n}(\mathbf{k})\tag{14}\\
&  =-\frac{e}{h}\sum_{n}^{\text{occu}}C_{n}.\nonumber
\end{align}
Thus when the chemical potential varies in the gap between $\varepsilon
_{1\mathbf{k}}$ and $\varepsilon_{2\mathbf{k}}$, only the lower band
$\varepsilon_{1\mathbf{k}}$ is occupied and $d\mathcal{M}/d\mu=-(e/h)C_{1}$.
For $\phi$=$\pi/3$, $C_{1}=-1$. Thus $d\mathcal{M}/d\mu=e/h$, i.e., the OM
linearly increases with the chemical potential in the insulating region I as
shown in Fig. 4(a). When the chemical potential touches the bottom of the
middle band, then the OM suddenly switches down and rapidly decreases again
when the chemical potential goes through the middle band. The turning behavior
at the band/gap contacts becomes divergent at $k_{B}T$=$0$. This discontinuity
is due to the singular behavior of $\Omega_{n}(\mathbf{k})$ and $m_{n}%
(\mathbf{k})$ at the BZ edge points $\mathbf{k}$=$\mathbf{k}_{0}$, which will
play their role when the $k$-integral is over the entire BZ. When the chemical
potential lies in the gap between the middle band and upper band, then the OM
goes up again and increases linearly with $\mu$ as shown in the insulating
region II in Fig. 4(a). Since the Chern number of middle band is zero, thus
from Eq. (14) and Fig. 4(a) one can see that the slope of the OM curve in the
insulating region II in Fig. 4(a) is the same as that in the insulating region I.

The totally different behavior of the OM in the metallic and insulating
regions, as shown in Fig. 4(a), is due to the different roles $M_{c}$ and
$M_{\Omega}$ play in these two regions. For further illustration, we show in
Fig. 4(b) $M_{c}$ (red curve) and $M_{\Omega}$ (blue curve) as a function of
the chemical potential, their sum gives $\mathcal{M}$ in Fig. 4(a). One can
see that overall $M_{c}$ and $M_{\Omega}$ have opposite contributions to
$\mathcal{M}$, which implies that these two parts carry opposite-circulating
currents. In each insulating area the conventional term $M_{c}$ keeps a
constant, which is due to the fact that the upper limit of the $k$-integral of
$m_{n}(\mathbf{k})$ is invariant as the chemical potential varies in the gap.
In the metallic region, however, since the occupied states varies with the
chemical potential, thus $M_{c}$ also varies with $\mu$, resulting in a
decreasing slope shown in Fig. 4(b). The Berry phase term $M_{\Omega}$ also
displays different behavior between insulating and metallic regions. In the
insulating region, $M_{\Omega}$ linearly increases with $\mu$, as is expected
from Eq. (3). In the metallic region, however, this term sensitively depends
on the topological property of the band in which the chemical potential is
located. For the lower and upper bands with nonzero Chen number, one can see
from Fig. 4(b) that $M_{\Omega}$ remains invariant, while for the middle band
of zero Chern number, it increases with the chemical potential $\mu$. On the
whole the comparison between Fig. 4(a) and Fig. 4(b) shows that the metallic
behavior of $\mathcal{M}$ is dominated by its conventional term $M_{c}$, while
in the insulating regime $M_{\Omega}$ plays a main role in determining the
behavior of $\mathcal{M}$.%

\begin{figure}[tbp]
\begin{center}
\includegraphics[width=1.0\linewidth]{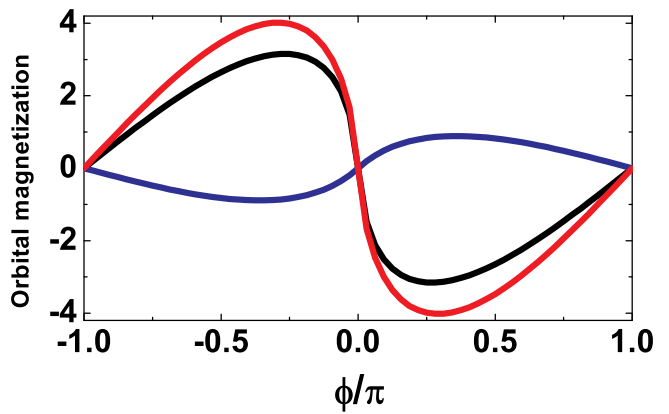}
\end{center}
\caption{(Color online). Orbital magnetization (black curve) of the \textit
{kagom\'{e}}
lattice, and its two components $M_{c}$ (red curve) and $M_{\Omega}$ (blue
curve) as a function of the spin chirality parameter $\phi
$ for the value of $\mu=-2$,
corresponding to the situation that the lower band is
partially occupied while the other two bands are empty.}
\label{fig5}
\end{figure}%
Figure 5 shows the OM $\mathcal{M}$ (black curve) and its two components
$M_{c}$ (red curve) and $M_{\Omega}$ (blue curve) as a function of spin
chirality parameter $\phi$ for the value of $\mu$=$-2$. This value of $\mu$
ensures that the lower band is partially occupied in the whole range of $\phi
$. Thus Figure 5 describes the metallic behavior of $\mathcal{M}$. Two
prominent features can be observed from Fig. 5: (i) The OM is antisymmetric
with respect to $\phi$, $\mathcal{M}(-\phi)$=$-\mathcal{M}(\phi)$. This
implies opposite circulating currents carried by Bloch states with $\phi$ and
those with $-\phi$; (ii) In the metallic situation, one can see that the
amplitude of $M_{\Omega}$ is much smaller than that of $M_{c}$. As a
consequence, the OM is dominated by its conventional part $M_{c}$ in the whole
range of $\phi$. In the insulating region (not shown in Fig. 5), however, the
amplitude of $M_{\Omega}$ largely increases and can be even larger than the
conventional contribution for some values of $\phi$ [see Fig. 4].

For transport studies of the OM, here we analyze the properties of ANE in the
2D \textit{kagom\'{e}} lattice. The relation between the OM and ANE has been
recently found \cite{Xiao2}. To discuss the transport measurement, it is
important to discount the contribution from the magnetization current, a point
which has attracted much discussion in the past. Cooper et al. \cite{Cooper}
have argued that the magnetization current cannot be measured by conventional
transport experiments. Xiao et al. \cite{Xiao2} have adopted this point and
built up a remarkable picture that the conventional orbital magnetic moment
$M_{c}$ does not contribute to the transport curent, while the Berry phase
term in Eq. (2) directly enters and therefore modifies the intrinsic transport
Hall current equation as follows%
\begin{equation}
\mathbf{j}_{\text{H}}\mathbf{=}-\frac{e^{2}}{\hbar}\mathbf{E}\times\sum
_{n}\int\frac{d^{2}k}{(2\pi)^{2}}f_{n}(\mathbf{r},\mathbf{k})\Omega
_{n}(\mathbf{k})\mathbf{-}\nabla\times\mathbf{M}_{\Omega}(\mathbf{r}),
\tag{15}%
\end{equation}
In the case of uniform temperature and chemical potential, obviously, the
second term is zero and the Hall effect of 2D \textit{kagom\'{e}} lattice is
featured by nonzero Chern number as discussed by Ohgushi et al. \cite{Ohgushi}
as well as in this paper. In the following, however, we turn to study another
situation, where the current-driving force is not provided by the electric
field ($\mathbf{E}$=0). Instead, it is provided by a statistical force, i.e.,
the gradient of temperature $T$. In this case, Eqs. (15) and (2) give the
expression of intrinsic thermoelectric Hall current as $j_{x}=\alpha
_{xy}(-\nabla_{y}T)$, where the anomalous Nernst conductivity $\alpha_{xy}$ is
given by \cite{Xiao2}\begin{widetext}
\begin{equation}
\alpha_{xy}=\frac{1}{T}\frac{e}{\hbar}\sum_{n}\int\frac{d^{2}k}{(2\pi)^{2}%
}\Omega_{n}\left[  \left(  \epsilon_{n\mathbf{k}}-\mu\right)  f_{n}+k_{B}%
T\ln\left(  1+e^{-\beta(\epsilon_{n\mathbf{k}}-\mu)}\right)  \right]  .
\tag{16}%
\end{equation}
\end{widetext}
%

\begin{figure}[tbp]
\begin{center}
\includegraphics[width=1.0\linewidth]{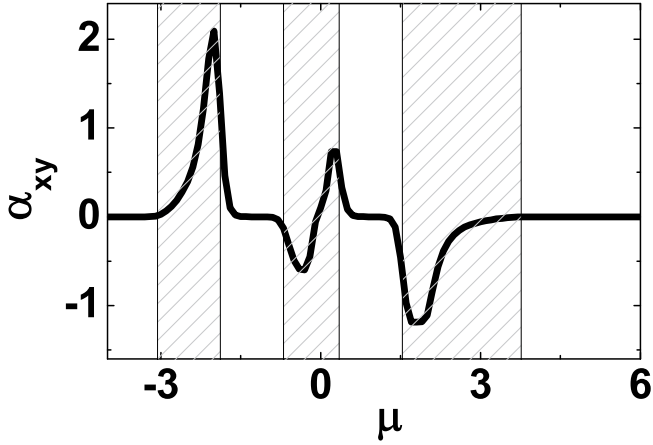}
\end{center}
\caption{The intrinsic anomalous Nernst conductivity $\alpha_{xy}%
$ (divided by the
temperature $T$) of the \textit{kagom\'{e}} lattice as a function of the
electron chemical potential $\mu$ for $\phi$=$\pi/3$ and $k_{B}T$=$0.05$. The
shaded areas correspond to the three groups of bands.}
\label{fig6}
\end{figure}%
Figure 6 shows $\alpha_{xy}$ of the 2D \textit{kagom\'{e}} lattice as a
function of the chemical potential for $\phi$=$\pi/3$ and $k_{B}T$=$0.05$. One
can see that the ANE disappears in the insulating regions, and when scanning
the chemical potential through the bands, there will appear peaks and valleys.
Remarkably, a similar peak-valley structure was also found by the recent
first-principles calculations in CuCr$_{2}$Se$_{4-x}$Br$_{x}$ compound
\cite{Xiao2}. The ANE of this compound was recently measured by Lee et al.
\cite{Lee} as a function of Br doping $x$ which is used to change the chemical
potential $\mu$. Due to the scarce data available, until now the peak-valley
structure of $\alpha_{xy}$ revealed in Fig. 6 and in Ref. \cite{Xiao2} has not
been found in experiment, and more direct experimental results are needed for
quantitative comparison with the theoretical results. Interestingly, the
expression for $\alpha_{xy}$ can be simplied at low temperature as the Mott
relation \cite{Xiao2},
\begin{equation}
\alpha_{xy}=-\frac{\pi^{2}}{3}\frac{k_{B}^{2}T}{e}\sigma_{xy}^{\prime}(\mu
_{0}),\tag{17}%
\end{equation}
where $\sigma_{xy}^{\prime}(\mu_{0})$ is the derivative of Hall conductivity
with respect to the zero-temperature chemical potential (Fermi energy)
$\mu_{0}$. Thus one can see that unlike AHE, ANE is given by the Fermi-surface
contribution of the band structure and Berry curvature. Another unique feature
of $\alpha_{xy}$ is its linear dependence of temperature.

Finally, let us consider the response of the OM to the external magnetic field
$B$. It should be noticed that the above semiclassical theory is carried out
up to the first order in the external perturbation \cite{Sund}. Thus our
discussion is valid in the weak magnetic field. From Eq. (2) one obtains the
magnetic susceptibility of the OM,
\begin{align}
\chi_{\text{o}}  &  =\frac{\partial\mathcal{M}}{\partial B}\tag{18}\\
&  =\sum_{n}\int\frac{d^{2}k}{(2\pi)^{2}}\left[  m_{n}^{2}(\mathbf{k}%
)\frac{\partial f_{n}}{\partial\mu}+\frac{e}{\hbar}\Omega_{n}(\mathbf{k}%
)m_{n}(\mathbf{k})f_{n}\right] \nonumber\\
&  \equiv\chi_{\text{o}}^{(c)}+\chi_{\text{o}}^{(\Omega)}.\nonumber
\end{align}
Obviously, the first conventional term $\chi_{\text{o}}^{(c)}$ describes the
Fermi-surface contribution to $\chi_{\text{o}}$ and thus will disappear in the
insulating region. Whereas the second term $\chi_{\text{o}}^{(\Omega)}$
denotes the Berry phase correction. Clearly, to calculate $\chi_{\text{o}%
}^{(\Omega)}$ all the occupied states within the Fermi level should be taken
into account. Therefore, the Berry phase correction will give a fundamental
contribution to $\chi_{\text{o}}$ in the insulating region for the systems
with nonzero Chern number.%

\begin{figure}[tbp]
\begin{center}
\includegraphics[width=1.0\linewidth]{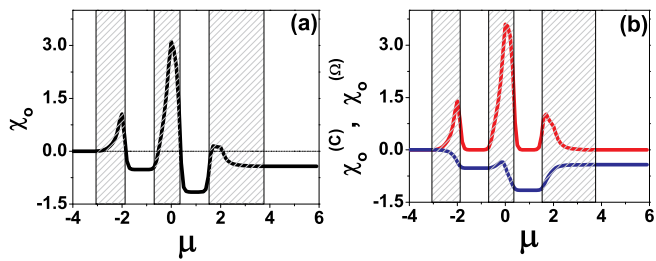}
\end{center}
\caption{(Color online). (a) Orbital magnetic susceptibility $\chi_{\text{o}%
}$ of the
\textit{kagom\'{e}} lattice, and (b) its two components contributed from
$M_{c}$ (red curve) and $M_{\Omega}$ (blue curve) as a function of the
electron chemical potential $\mu$ for $\phi$=$\pi/3$ and $k_{B}T$=$0.05$.}
\label{fig7}
\end{figure}%
Figures 7(a)-(b) show the orbital magnetic susceptibility and its two
components as a function of the chemical potential, respectively. From Fig.
7(b) one can see that the conventional part $\chi_{\text{o}}^{(c)}$ always
displays the paramagnetic property, while the Berry phase correction
contributes a diamagnetic response to $B$. Again, in the metallic regions, the
total magnetic susceptibility is dominated by its conventional part and thus
displays the paramagnetic feature. Whereas in the insulating regions the
nonzero $\chi_{\text{o}}$ comes solely from the Berry phase correction and
displays the diamagnetic feature.

Up to now, we have concentrated on the ferromagnet represented by the
double-exchange model. As Ohgushi et al. \cite{Ohgushi} have pointed out, the
present theory is also applicable to the ferromagnet based on the Hubbard
model. Furthermore, the spin-orbit coupling gives the spin anisotropies, which
introduces the tilting of the spins from the perfect ferromagnetic alignment
as assumed in Eq. (4). The present 2D \textit{kagom\'{e}} lattice may be
relevant to the recent experiments on pyrochlore compounds R$_{2}$Mo$_{2}%
$O$_{7}$ $($R=Nd, Sm, Gd$)$ \cite{Taguchi, Katsufuji}, which are itinerant
ferromagnets on the verge of a Mott transition on the pyrochlore lattice. It
can be expected that easy-axis spin anisotropy in these compounds produces the
spin chirality by the symmetry consideration. Thus the present results of the
OM properties may be verified in such systems.

\section{Conclusion}

In summary, we have theoretically studied the properties of OM and its effects
in the 2D spin-chiral ferromagnetic \textit{kagom\'{e}} lattice. The spin
chirality parameter $\phi$ \ in the lattice produces nonzero Chern number and
results in profound effects on the OM properties. We have found that the two
parts $M_{c}$ and $M_{\Omega}$ in OM opposite each other, and yield the
paramagnetic and diamagnetic responses respectively. In particular, due to its
Chern-number property, the magnetic susceptibility of $M_{\Omega}$ remains to
be a nonzero constant when the Fermi energy is located in the energy gap of
the \textit{kagom\'{e}} lattice. It has been further shown that the OM
displays fully different behavior in the metallic and insulating regions,
because of different roles $M_{c}$ and $M_{\Omega}$ play in these two regions.
The anomalous Nernst conductivity has also been studied, which displays a
peak-valley structure as a function of the electron chemical potential. The
experiments on ferromagnets are urgently expected to realize these interesting
theoretical results.

\begin{acknowledgments}
This work was supported by CNSF under Grant No. 10544004 and 10604010.
\end{acknowledgments}

\end{document}